\documentclass[sigconf,natbib=true]{acmart}
\usepackage{orcidlink}

\usepackage[utf8]{inputenc}
\usepackage{booktabs,multirow}
\usepackage{balance}
\usepackage{enumitem}
\usepackage{adjustbox}
\usepackage{color}
\usepackage{rotating}
\usepackage{amsmath}
\usepackage{pifont}
\usepackage{array}
\usepackage{bbm}
\usepackage{multirow}
\usepackage{colortbl} 
\usepackage[normalem]{ulem}
\useunder{\uline}{\ul}{}
\usepackage{graphicx}
\usepackage{subcaption} 
\usepackage{kotex} % 한글 사용
\usepackage[linesnumbered,algoruled,boxed,lined]{algorithm2e}

\usepackage{booktabs}

\newcommand{\ie}{{\it i.e.}}

\newcommand{\argmax}{\operatornamewithlimits{argmax}}

\AtBeginDocument{%
  }
    
\copyrightyear{2025}
\acmYear{2025}
\setcopyright{acmlicensed}
\acmConference[CIKM '25] {Proceedings of the 34th ACM International Conference on Information and Knowledge Management}{November 10--14, 2025}{Seoul, Republic of Korea.}
\acmBooktitle{Proceedings of the 34th ACM International Conference on Information and Knowledge Management (CIKM '25), November 10--14, 2025, Seoul, Republic of Korea}
\acmISBN{979-8-4007-2040-6/2025/11}
\acmDOI{10.1145/3746252.3761190}

\begin{document}

\title{MUFFIN: Mixture of User-Adaptive Frequency Filtering for Sequential Recommendation}

\author{Ilwoong Baek} \authornote{Both authors contributed equally to this research.}
\orcid{0009-0000-6515-237X}
\affiliation{
  \institution{Sungkyunkwan University}
  \city{Suwon}
  \country{Republic of Korea}}
\email{alltun100@skku.edu}

\author{Mincheol Yoon} \authornotemark[1]
\orcid{0009-0003-0697-915X}
\affiliation{
  \institution{Sungkyunkwan University}
  \city{Suwon}
  \country{Republic of Korea}}
\email{yoon56@skku.edu}

\author{Seongmin Park} 
\orcid{0009-0004-8211-1691}
\affiliation{
  \institution{Sungkyunkwan University}
  \city{Suwon}
  \country{Republic of Korea}}
\email{psm1206@skku.edu}

\author{Jongwuk Lee}\authornote{Corresponding author}
\orcid{0000-0001-9213-7706}
\affiliation{
  \institution{Sungkyunkwan University}
  \city{Suwon}
  \country{Republic of Korea}}
\email{jongwuklee@skku.edu}

\begin{abstract}\label{sec:abstract}

Sequential recommendation (SR) aims to predict users’ subsequent interactions by modeling their sequential behaviors. Recent studies have explored \emph{frequency domain analysis}, which effectively models periodic patterns in user sequences. However, existing frequency-domain SR models still face two major drawbacks: (i) \emph{limited frequency band coverage}, often missing critical behavioral patterns in a specific frequency range, and (ii) \emph{lack of personalized frequency filtering}, as they apply an identical filter for all users regardless of their distinct frequency characteristics. To address these challenges, we propose a novel frequency-domain model, \emph{\textbf{M}ixture of \textbf{U}ser-adaptive \textbf{F}requency \textbf{FI}lteri\textbf{N}g (\textbf{MUFFIN})}, operating through two complementary modules. (i) The \emph{global filtering module (GFM)} handles the entire frequency spectrum to capture comprehensive behavioral patterns. (ii) The \emph{local filtering module (LFM)} selectively emphasizes important frequency bands without excluding information from other ranges. (iii) In both modules, the \emph{user-adaptive filter (UAF)} is adopted to generate user-specific frequency filters tailored to individual unique characteristics. Finally, by aggregating both modules, MUFFIN captures diverse user behavioral patterns across the full frequency spectrum. Extensive experiments show that MUFFIN consistently outperforms state-of-the-art frequency-domain SR models over five benchmark datasets. The source code is available at \url{https://github.com/ilwoong100/MUFFIN}.

\end{abstract}
\maketitle

\section{Introduction}\label{sec:introduciton}

\begin{figure}
\centering
\includegraphics[width=1.0\linewidth]{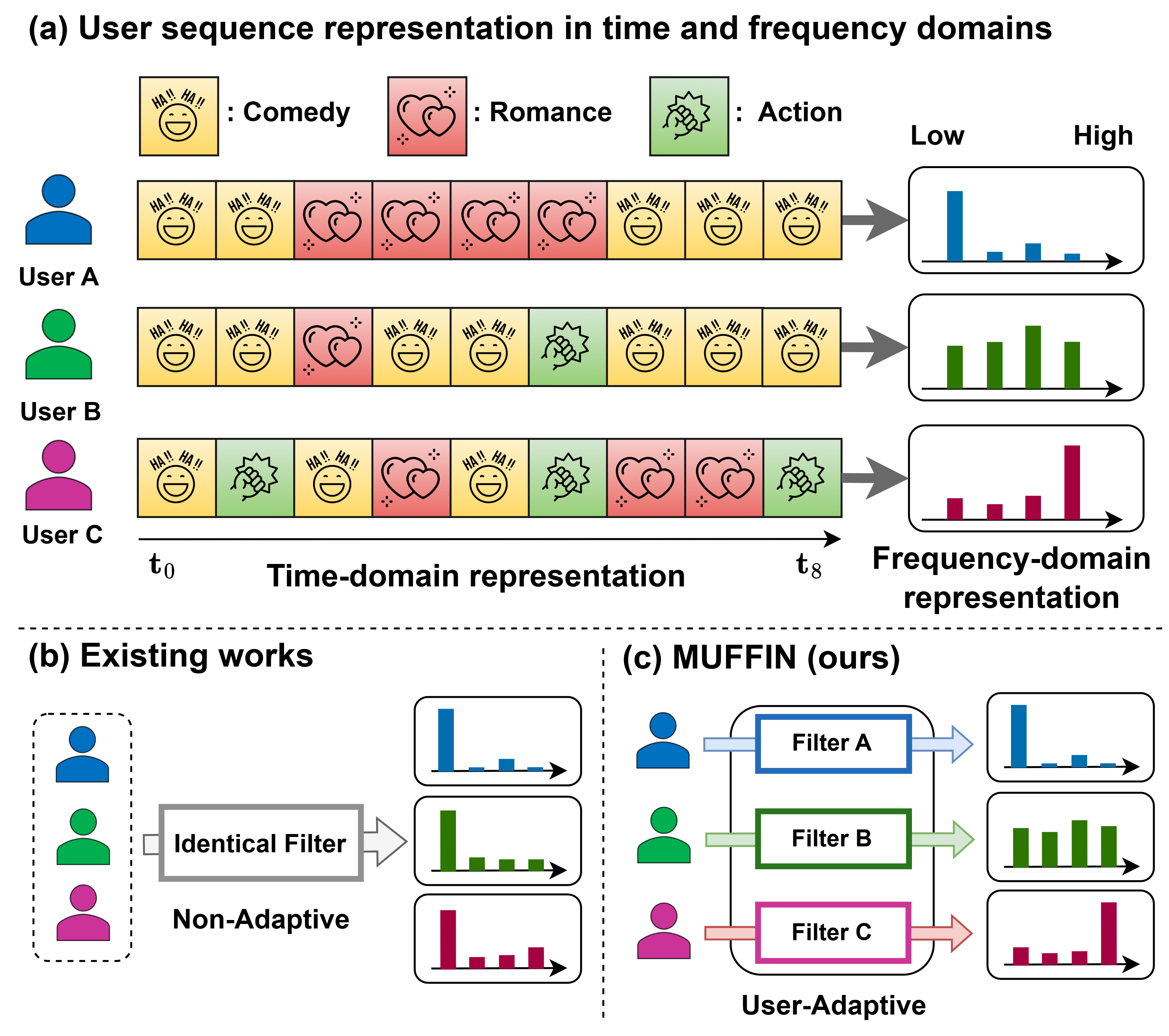} \\

\vspace{-2mm}
\caption{ (a) Illustration of user sequences converted to frequency domain. Each user exhibits distinct frequency characteristics. (b) Existing works apply an identical filter to users, which fails to capture individual frequency characteristics. (c) Our model performs user-adaptive frequency filtering tailored to individual user characteristics.
}

\label{fig:fig1}

\vspace{-2mm}
\end{figure}

Sequential recommendation (SR)~\cite{fang2020deep, FangZSG20SeqSurvey, WangHWCSO19SeqSurvey} aims to predict users’ next interactions by modeling their historical behavior sequences. Unlike traditional recommendation settings that treat user-item interactions as independent instances, SR models~\cite{SunLWPLOJ19BERT4Rec, KangM18SASRec, zhou2020s3, XieSLWGZDC22CL4SRec, TangW18caser, aaai/Shin0WP24BSARec, park2025temporal} focus on capturing temporal dynamics and evolving user preferences in user sequences. These user sequences often exhibit intricate patterns at multiple levels of granularity, ranging from high-level transitions in user interests to subtle contextual variations~\cite{wangsequential,gao2023survey, wu2022survey}. Therefore, the core challenge of SR lies in effectively modeling diverse patterns in user interactions over time.

Recent advances in deep learning have significantly enhanced the SR performance, particularly by adopting transformer-based models using self-attention mechanisms. These models~\cite{SunLWPLOJ19BERT4Rec, KangM18SASRec} excel at modeling long-range dependencies within user sequences, effectively identifying item-to-item relationships. Despite their strengths, recent studies~\cite{aaai/Shin0WP24BSARec, park2022vision} have highlighted a key limitation: self-attention mechanisms often struggle to capture fine-grained behavioral patterns, thereby impeding their ability to fully model the diverse and dynamic nature of user behavior.

To address this limitation, an emerging line of research~\cite{ZhouYZW22FMLPRec, aaai/Shin0WP24BSARec, DuYZQZ0LS23FEARec, DuYZF0LS023SLIME4Rec, xiao2024tfcsrec} has explored \emph{frequency-domain representation}. It is motivated by the observation that user sequences often exhibit inherent periodicities and multi-scale variations. Using the discrete Fourier transform (DFT), a time-domain representation is transformed into the frequency-domain representation. Therefore, complex behavioral patterns are decomposed into interpretable spectral components, where low-frequency signals correspond to gradual shifts in user interests, and high-frequency signals capture abrupt or short-term interest changes~\cite{sejdic2011fractional}.

As illustrated in Figure~\ref{fig:fig1}(a), three users show distinct frequency characteristics based on their behavioral patterns. User A (blue) shows gradual transitions (\ie, two changes) with dominant low-frequency components, and User B (green) shows a multi-band spectrum with both strong low- and high-frequency components, as comedy films dominate their consumption while intermittently exploring other genres (\ie, four changes). Meanwhile, User C (pink) exhibits rapid genre switching with frequent item transitions (\ie, seven changes), resulting in a frequency spectrum dominated by high-frequency components.

Although existing frequency-domain SR models~\cite{ZhouYZW22FMLPRec, DuYZF0LS023SLIME4Rec, aaai/Shin0WP24BSARec} attempt to capture diverse user interaction patterns, they still suffer from two limitations. (i) \emph{Limited frequency band coverage}: existing models face a trade-off between full frequency spectrum analysis and detailed frequency band-specific modeling. As observed in Figure~\ref{fig:fig1}(a), it is necessary to integrate global spectrum understanding and local band-specific analysis simultaneously. However, existing models cannot achieve this dual capability. (ii) \emph{Lack of personalized frequency filtering}: existing models apply identical frequency filters regardless of the different frequency characteristics of users. As shown in Figure~\ref{fig:fig1}(b), when a low-frequency-emphasized filter is equally applied to three users, capturing each user's unique behavioral pattern is difficult, leading to information dilution.

To address this challenge, we propose \emph{Mixture of \textbf{U}ser-adaptive \textbf{F}requency \textbf{FI}lteri\textbf{N}g (\textbf{MUFFIN})}, which operates through two complementary modules. The \emph{global filtering module (GFM)} processes the entire frequency spectrum simultaneously to capture comprehensive user characteristics across all frequency bands, capturing the users' overall behavioral tendencies. In contrast, the \emph{local filtering module (LFM)} divides the spectrum into distinct frequency bands and selectively emphasizes the important ones while preserving information from the others. To enhance user adaptiveness for each module, we also introduce a \emph{user-adaptive filter (UAF)} that dynamically generates personalized filters tailored to each user's unique characteristics. Finally, MUFFIN ensures personalized filtering that adapts to diverse user behavior patterns, enabling both modules to adjust their filtering strength based on individual user patterns.
As depicted in Figure~\ref{fig:fig1}(c), MUFFIN generates user-adaptive filters and effectively identifies individual behavioral patterns. Through extensive experiments, MUFFIN  is compared against eight SR models and outperforms state-of-the-art frequency-domain models over five benchmark datasets.

The key contributions of this paper are summarized as follows:
\begin{itemize}[leftmargin=5mm, topsep=1pt]
    \item We propose a novel dual filtering architecture consisting of the \emph{global filtering module (GFM)} and the \emph{local filtering module (LFM)}. The GFM processes the entire frequency spectrum to capture comprehensive behavioral patterns across all frequency ranges, while the LFM selectively emphasizes specific frequency bands to model user-specific behavioral patterns without losing any frequency information.
    \vspace{1mm}
    
    \item We employ the \emph{user-adaptive filter (UAF)} that serves as a core mechanism. It enables both GFM and LFM to perform user-specific filtering by generating personalized filters based on individual frequency-domain representations.
    \vspace{1mm}
    
    \item Through extensive experiments on five datasets over eight SR models, we demonstrate that MUFFIN achieves over state-of-the-art models, showcasing its ability to adapt to user-specific patterns and capture diverse behavior preferences.
\end{itemize}

\section{Preliminaries}\label{sec:preliminaries}
% outline
% 1. Sequential Recommendation problem statement
% 2. DFT and FFT

\subsection{Problem Formulation}
% The goal of SR is to predict the next item a user will interact with based on user's historical behavior.

Let $\mathcal{U}$ and $\mathcal{I}$ denote the sets of users and items, respectively, where $|\mathcal{U}|$ and $|\mathcal{I}|$ are the total number of users and items. For each user $u \in \mathcal{U}$, we are given a user  sequence $S_u=[i_1, i_2, ..., i_{|S_u|}]$, where $i_j \in \mathcal{I}$ indicates the $j$-th item in the user sequence, and $|S_u|$ is the length of the user sequence.
Given a user sequence $S_u$, the goal of SR is to predict the next item $i_{|S_u|+1}$ with which user $u$ will most likely interact. It is formulated as training an optimal model parameter $\theta^*$ that maximizes the conditional probability:
\begin{equation}
    \theta^* = \argmax_\theta P_\theta(i = i_{|S_u|+1} \ | \ S_u).
\end{equation}

For each prediction, the SR model parameterized by $\theta$ produces a probability distribution over the item set $\mathcal{I}$. Thus, higher probabilities indicate items the user is more likely to interact with next.

\subsection{Fourier Transform}\label{fouriertransform}
The discrete Fourier transform (DFT) is a fundamental component of digital signal processing~\cite{rabiner1975theory, soliman1990continuous}, converting a sequence in the time domain into the frequency domain. Given an input sequence $\mathbf{S} \in \mathbb{R}^{N}$ with length $N$, the DFT is denoted as $\mathcal{F}: \mathbb{R}^{N} \to \mathbb{C}^{N}$, and its inverse, \ie, the inverse discrete Fourier transform (IDFT), is denoted as $\mathcal{F}^{-1}: \mathbb{C}^{N} \to \mathbb{R}^{N}$. The DFT can be performed as:
\begin{equation} \label{eq:dft}
    \mathbf{F} = \mathcal{F}(\mathbf{S}) = \frac{1}{\sqrt{N}}
    \begin{bmatrix}
    1 & 1 & \cdots & 1 \\
    1 & e^{\frac{-2\pi i}{N}} & \cdots & e^{\frac{-2\pi i(N-1)}{N}} \\
    \vdots & \vdots & \ddots & \vdots \\
    1 & e^{\frac{-2\pi i(N-1)}{N}} & \cdots & e^{\frac{-2\pi i(N-1)^{2}}{N}}
    \end{bmatrix} \mathbf{S},
\end{equation}
where $i$ is the imaginary unit, and $\mathbf{F} \in \mathbb{C}^{N}$ is the frequency component of the sequence $\mathbf{S}$.
In practice, we utilize the Fast Fourier Transform (FFT)~\cite{gabor1946theory} algorithm, which efficiently computes the DFT with a computational complexity of $O(N\log N)$ compared to the direct DFT computation of $O(N^2)$. Furthermore, since our sequence $\mathbf{S}$ consists of real numbers, its Fourier transform exhibits conjugate symmetry:
\begin{equation}
\mathbf{F}[k] = \mathbf{F}[N-k]^* ,  k = 1,2,..., \lfloor N/2⌋ \rfloor,
\end{equation}
where $^*$ denotes the complex conjugate. Due to this symmetry property, we use the real-valued fast Fourier transform (RFFT)~\cite{nussbaumer1982fast, sorensen1987real}, which computes only the non-redundant half of the spectrum. This reduces the output dimension from $N$ to $\lfloor \frac{N}{2} \rfloor + 1$ while preserving all unique frequency information. Using RFFT, the frequency-domain representation is:
\begin{equation}
    \mathbf{F} = \mathcal{F}_R(\mathbf{S}) \in \mathbb{C}^{M}, \text{ where } M = \left\lfloor\frac{N}{2}\right\rfloor + 1,
\end{equation}
where $\mathcal{F}_R(\cdot)$ denotes RFFT. This reduces computational overhead and maintains all essential frequency components needed for SR. For notational convenience, we denote the RFFT $\mathcal{F}_R$ as $\mathcal{F}$.

To capture the frequency band in $\mathbf{F}$, we can extract the frequency components from indices $f_t$ to $f_{t+1}$, where $f_t$ denotes the starting index of the $t$-th frequency band. It represents the particular frequency band in the transformed sequence.
\begin{equation} \label{eq:band}
    \mathbf{B}_t = \mathbf{F}[f_t:f_{t+1}], \ \ 
\end{equation}
where $\mathbf{B}_t \in \mathbb{C}^{f_{t+1}-f_t}$ and $[:]$ indicates the slice operation.
In $\mathbf{B}_{t}$, smaller values of ${t}$ correspond to low frequencies, while larger values of ${t}$ correspond to high frequencies.

% \subsection{Frequency Filter Layer}\label{ffilterlayer}
% frequency에서의 filter layer는 입력 시퀀스에 대한 주파수 성분의 중요도를 조절하는 핵심 구성 요소이다. 이는 신호 처리 분야에서 널리 사용되는 주파수 필터링 개념을 기반으로 한다. 빈도 도메인 필터 레이어는 다음과 같이 정의된다: 
% \begin{equation} \label{eq:filterL}
%     f(\mathbf{S}) = \mathcal{F}^{-1}(\mathbf{W} \odot \mathcal{F}(\mathbf{S}))
% \end{equation}
% 여기서 $\mathbf{S} \in \mathbb{R}^{n \times d}$는 입력 시퀀스 표현, $\mathcal{F}$는 푸리에 변환, $\mathcal{F}^{-1}$는 역 푸리에 변환을 나타낸다. $\mathbf{W} \in \mathbb{C}^{m \times d}$는 주파수 도메인에서의 필터 가중치이며, $\odot$는 요소별 곱셈(element-wise multiplication)을 의미한다. 필터 가중치 $\mathbf{W}$는 각 주파수 성분의 중요도를 결정하며, 특정 주파수 대역을 강조하거나 억제하는 역할을 한다.

% 이러한 필터링 메커니즘은 사용자 행동 패턴과 관련된 중요한 주파수 성분을 강조하여 추천 성능을 향상시킨다. 기존의 빈도 도메인 SR 모델들은 모든 사용자에게 동일하게 적용되는 공유 필터 가중치 $\mathbf{W}$를 학습한다. 그러나 이 접근법은 사용자 간의 행동 패턴 차이를 고려하지 못하는 근본적인 한계를 가진다. 본 논문에서 제안하는 MUFFIN은 각 사용자의 고유한 빈도 스펙트럼 특성에 적응하는 사용자 적응형 빈도 필터(UAF)를 도입하여 이러한 한계를 극복한다.

\subsection{Learnable Frequency Filter}\label{sec:ffilterlayer}
The filtering layer in the frequency domain is a key component that adjusts the importance of frequency components for input sequences. This is based on the concept of frequency filtering widely used in signal processing, and some studies~\cite{ZhouYZW22FMLPRec, DuYZF0LS023SLIME4Rec} attempted to implement it using learnable filters in SR. The learnable frequency filter layer is defined as follows:
\begin{equation} \label{eq:filterL}
    f(\mathbf{S}) = \mathcal{F}^{-1}(\mathcal{F}\left( \mathbf{S}) \odot \mathbf{W} \right),
\end{equation}
where $\mathcal{F}^{-1}$ represents inverse RFFT, $\mathbf{W} \in \mathbb{C}^{M}$ is learnable frequency filter, and $\odot$ denotes element-wise multiplication.

This filter $\mathbf{W}$ is optimized during model training. In other words, it captures specific patterns in the sequence by emphasizing or suppressing certain frequency bands.

% This filtering mechanism enhances recommendation performance by emphasizing important frequency components related to user behavior patterns.

% 주파수 도메인에서의 필터링 레이어는 입력 시퀀스에 대한 주파수 성분의 중요도를 조절하는 핵심 구성 요소이다. 이는 신호 처리 분야에서 널리 사용되는 주파수 필터링 개념을 기반으로 하며, 순차적 추천 시스템에서는 학습 가능한 필터를 통해 구현된다. 학습 가능한 주파수 필터링 레이어는 다음과 같이 정의된다:
% \begin{equation} \label{eq:filterL}
%     f(\mathbf{S}) = \mathcal{F}^{-1}(\mathbf{W} \odot \mathcal{F}(\mathbf{S}))
% \end{equation}
% 여기서 $\mathbf{W} \in \mathbb{C}^{M}$는 학습 가능한 복소 필터 가중치이며, $\odot$는 요소별 곱셈(element-wise multiplication)을 의미한다.
% 학습 가능한 필터 $\mathbf{W}$는 모델 훈련 과정에서 최적화되며, 각 주파수 성분의 중요도를 결정한다.\cite{ZhouYZW22FMLPRec} 구체적으로, 이 필터는 특정 주파수 대역을 강조하거나 억제하여 시퀀스의 중요한 패턴을 포착하는 역할을 한다. 이러한 필터링 메커니즘은 사용자 행동 패턴과 관련된 중요한 주파수 성분을 강조하여 추천 성능을 향상시킨다.

% 기존의 주파수 도메인 SR 모델들(예: FMLP-Rec\cite{zhou2022filter}, SLIME4Rec\cite{DuYZF0LS023SLIME4Rec})은 모든 사용자에게 동일하게 적용되는 공유 필터 가중치 $\mathbf{W}$를 학습한다. 이러한 균일한 필터링 접근법은 데이터셋 내 지배적인 사용자들의 패턴에 크게 영향을 받을 수 있으며, 개별 사용자의 고유한 행동 특성을 반영하지 못하는 근본적인 한계를 가진다. 또한, 모든 주파수 대역을 동시에 처리하거나 특정 레이어에서 제한된 주파수 대역만 처리하는 방식은 다양한 사용자 행동 패턴을 포괄적으로 모델링하는 데 제한이 있다.

% \input{sec-study}
% \input{sec-analysis}
\begin{figure*}
\centering
\includegraphics[width=0.85\linewidth]{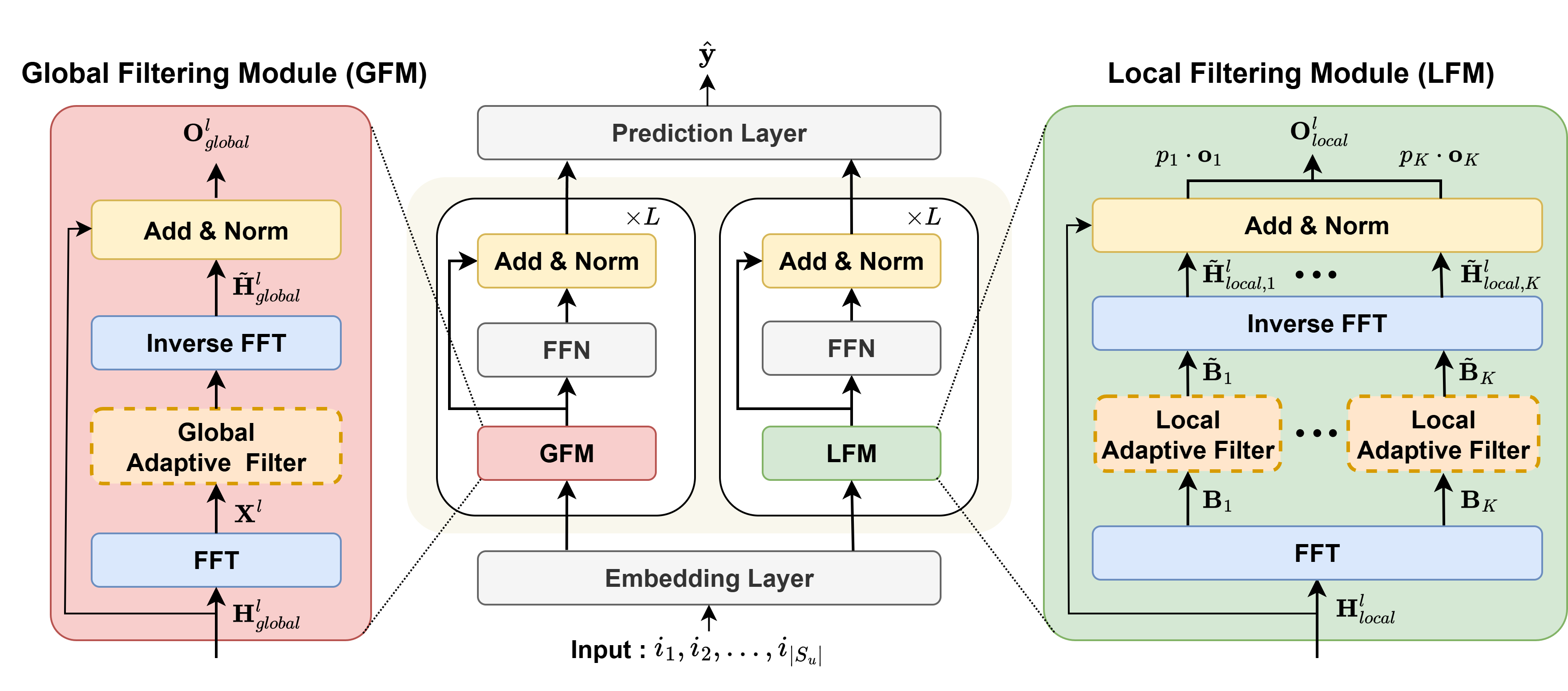}
\vspace{-3mm}
\caption{Overview of MUFFIN, which trains with two parallel modules: global filtering module (GFM) and local filtering module (LFM). Both modules utilize the user-adaptive filter (UAF) to adjust to individual user behavior patterns in the frequency domain. A mixture of the module outputs is used to predict the target item. }\label{fig:model}
\vspace{-2mm}

\end{figure*}
%  Auxiliary loss that predicts the target item with each module's output is also utilized for optimization.

\section{Proposed Model: MUFFIN}\label{sec:methods}

In this section, we present \emph{\textbf{M}ixture of \textbf{U}ser-adaptive \textbf{F}requency \textbf{FI}lteri\textbf{N}g (\textbf{MUFFIN})}, capturing diverse user behavior patterns. As depicted in Figure~\ref{fig:model}, it comprises two complementary filtering modules: \emph{global filtering module (GFM)} and \emph{local filtering module (LFM)}, each designed to extract different aspects of user behavior. To enable user adaptiveness in both modules, we also incorporate a \emph{user-adaptive filter (UAF)}, which dynamically generates personalized filters tailored to each user’s frequency-domain characteristics. Section~\ref{sec:filtering} details these key components of MUFFIN, and Section~\ref{sec:training} describes the training procedure of MUFFIN.

%In this section, we present \textbf{M}ixture of \textbf{U}ser-adaptive \textbf{F}requency \textbf{FI}lteri\textbf{N}g (\textbf{MUFFIN}). We first explain the key components of MUFFIN (Section~\ref{sec:filtering}). As illustrated in Figure~\ref{fig:model}, MUFFIN consists of dual filtering modules designed to capture different aspects of user behavior patterns: \emph{global filtering module (GFM)} and \emph{local filtering module (LFM)}. To achieve user adaptiveness in both modules, we also adopt a \emph{user-adaptive frequency filter (UAF)} that dynamically generates personalized filters based on each user's unique frequency-domain representation. We then explain the training process of MUFFIN to capture diverse sequence patterns (Section~\ref{sec:training}).

\subsection{Mixture of Filtering Modules}\label{sec:filtering}

\vspace{1mm}
\noindent
\textbf{Embedding layer}.
The item embedding matrix \( \mathbf{E} \in \mathbb{R}^{|\mathcal{I}| \times d} \), where \( d \) is the embedding dimension, is employed to project items into a latent embedding space. Given a user sequence \( S_u = [i_1, i_2, \dots, i_{|S_u|}] \), each item $i_j$ is represented as its corresponding embedding vector $\mathbf{e}_j \in \mathbb{R}^{d}$. These item embeddings are concatenated to form the initial sequence representations \(\mathbf{H}^0 = [\mathbf{e}_1, \mathbf{e}_2, \dots, \mathbf{e}_{|S_u|} ]\), where \(\mathbf{H}^0 \in \mathbb{R}^{n \times d}\), and $n$ is the maximum length of the sequence.

\vspace{1mm}
\noindent
\textbf{User-adaptive filter (UAF)}. It is dynamically generated to enable personalized filtering within global and local filtering modules. The initial sequence representation $\mathbf{H}^{0}$ is transformed from the time domain into the frequency domain by the Fourier transform:
\begin{equation}
   \mathbf{X}^{0} = \mathcal{F}(\mathbf{H}^{0}) \in \mathbb{C}^{m \times d},
\end{equation}

\noindent
where \(m= \lfloor n/2 \rfloor+1\), and $\mathbf{X}^0$ denotes  the initial representation in the frequency domain.

Inspired by recent advances in frequency-based filtering for computer vision~\cite{lin2023deep}, we design a convolution filter ${h}(\cdot)$ to capture user-specific behavior patterns:
\begin{equation}
    \mathbf{C}(\mathbf{X}^{0}) = \text{BatchNorm1D}\left(\texttt{Conv1D}(\|\mathbf{X}^{0}\|^{\top}) \right) \in \mathbb{R}^{d \times m},
\end{equation}
\begin{equation}\label{eq:hfunction}
    h(\mathbf{X}^{0}) = \sigma \left(\mathbf{C}(\mathbf{X}^{0}) \right)^{\top}\in \mathbb{R}^{m \times d},
\end{equation} 
where $\sigma$ denotes sigmoid function, and $\|\cdot\|$ denotes the amplitude (\ie, magnitude) of the complex-valued tensor. \(\top\) indicates the matrix transpose operator. The $\texttt{Conv1D}$~\cite{kiranyaz20211d} operation employs a convolutional kernel \(\mathbf{W}_{c} \in \mathbb{R}^{d \times d \times c}\), where \(c\) is a hyperparameter for kernel size. By leveraging amplitude information and applying 1D convolution, we compute a weighted aggregation of each frequency component and its neighboring frequencies. Therefore, this convolutional filter captures meaningful patterns in the frequency-amplitude domain to construct personalized representations.

%By utilizing the amplitude information and performing the 1D convolution, it calculates the weighted sum of each frequency component and its neighboring components. This allows the model to dynamically adjust the filter strength required for each user.
% UAF is then passed to the dual filtering module for utilization.

% We propose two complementary filtering modules (\ie, GFM and LFM) to cover all frequency bands.
% Each module can capture diverse behavioral patterns within user sequences from different perspectives.

\vspace{1mm}
\noindent
\textbf{Global filtering module (GFM).}
It captures high-level behavioral patterns by analyzing the full frequency spectrum of user sequences. At the $l$-th layer, the frequency-domain representation $\mathbf{X}^{l}_{global}$ is computed as:
\begin{equation}
    \mathbf{X}^{l}_{global} = \mathcal{F}(\mathbf{H}^{l}_{global}),
\end{equation}
where $\mathbf{H}^{0}_{global}$ corresponds to the initial representation $\mathbf{H}^{0}$.

To modulate frequency components across the entire spectrum, we adopt a learnable filter \(\mathbf{W}_{global} \in \mathbb{C}^{m \times d}\). It is refined using the user-adaptive filter (UAF) as follows: 
\begin{equation}
    \tilde{\mathbf{W}}^{l}_{global} = h(\mathbf{X}^{0}) \odot \mathbf{W}^{l}_{global},
\end{equation}

\noindent
where $h(\cdot)$ denotes the amplitude-based convolutional filter described in Eq.~\eqref{eq:hfunction}. Notably, the UAF takes only the initial frequency representation \(\mathbf{X}^{0}\) as input, rather than the layer-specific input $\mathbf{X}^{l}$ for $l > 0$. This enables the filter to encode user-specific characteristics from the original sequence spectrum. 

The global adaptive filter \(\tilde{\mathbf{W}}^{l}_{global}\) is then applied to scale the frequency-domain representation. It is transformed back to the time domain via the inverse Fourier transform:
\begin{equation}
    \tilde{\mathbf{H}}^{l}_{global} = \mathcal{F}^{-1}(\mathbf{X}^{l}_{global} \odot \tilde{\mathbf{W}}^{l}_{global}),
\end{equation}

\noindent
where $\tilde{\mathbf{H}}^{l}_{global} \in \mathbb{R}^{n \times d}$. To ensure stable training and effective gradient propagation~\cite{he2016deep}, the final output of the GFM incorporates residual connections, dropout, and layer normalization:
\begin{equation}
    \mathbf{O}_{global}^{l} = \text{LayerNorm}\left(\mathbf{H}^{l}_{global} + \text{Dropout}(\tilde{\mathbf{H}}^{l}_{global})\right).
\end{equation}

\vspace{1mm}
\noindent
\textbf{Local filtering module (LFM).} It is designed to capture fine-grained patterns by partitioning the frequency spectrum and emphasizing specific frequency bands. At the $l$-th layer, the frequency-domain representation of the user sequence \(\mathbf{X}^{l}_{local}\) is obtained by applying the Fourier transform to the input sequence \(\mathbf{H}^{l}_{local}\), and dividing the result into \(K\) contiguous frequency bands:
\begin{equation} \label{eq:local_divide}
    \mathbf{X}^{l}_{local} = [\mathbf{B}_{1}^{l}; \mathbf{B}_{2}^{l}; ...; \mathbf{B}_{K}^{l}] = \mathcal{F}(\mathbf{H}^{l}_{local}), %\odot \mathbf{W}^{l}_{global},
\end{equation}
where each \(\mathbf{B}_{t}^{l}=\mathbf{X}^{l}_{local}[f_{t}:f_{t+1}] \in \mathbb{C}^{a_t \times d}\) denotes the $t$-th frequency band with starting index $f_1 = 0$, and $K$ is a hyperparameter to control the number of frequency bands. The band size $a_t$ determines the number of frequency components assigned to each band \footnote{Here, $a_t = \lfloor \frac{t \times m}{K} \rfloor - \lfloor \frac{(t-1) \times m}{K} \rfloor$ and $\sum_{t=1}^K a_t = m$, ensuring the complete coverage of the frequency spectrum.}. This band-wise decomposition enables our model to focus on local frequency patterns, facilitating the extraction of subtle behavioral signals that may not be captured by global representations.

Similar to GFM, LFM also utilizes a learnable filter $\mathbf{W}_{local} \in \mathbb{C}^{m \times d}$ to control the contribution of each frequency component. To incorporate user-specific behavior, we apply UAF to generate a local adaptive filter at the $l$-th layer:
\begin{equation}
    \tilde{\mathbf{W}}^{l}_{local} = h(\mathbf{X}^{0}) \odot \mathbf{W}^{l}_{local}.
\end{equation}

This filter is then divided into $K$ bands to match the band-wise structure of the frequency representation:
\begin{equation}
    \tilde{\mathbf{W}}^{l}_{local} = [\tilde{\mathbf{W}}^{l}_{local,1};  \tilde{\mathbf{W}}^{l}_{local,2}; \dots; \tilde{\mathbf{W}}^{l}_{local,K}],
\end{equation}
where $\tilde{\mathbf{W}}^{l}_{local,t} \in \mathbb{C}^{a_t \times d}$ is applied to the $t$-th frequency band. 

To preserve the full frequency resolution required for the inverse Fourier transform, we apply zero-padding after filtering each frequency band:
\begin{align}
     \tilde{\mathbf{B}}^{l}_{t} &= \text{ZeroPadding}(\mathbf{B}^{l}_{t} \odot\tilde{\mathbf{W}}^{l}_{local,t}) \nonumber \\
    &= [\mathbf{0}_{f_t \times d}; \mathbf{B}^{l}_{t} \odot\tilde{\mathbf{W}}^{l}_{local,t}; \mathbf{0}_{(m-f_{t+1}) \times d}]\in \mathbb{C}^{m\times d},
\end{align}

\noindent
where $\mathbf{0}_{f_t \times d}$ and $\mathbf{0}_{(m - f_{t+1}) \times d}$ are zero matrices that pad the beginning and end of the spectrum, respectively.

Each zero-padded frequency representation is transformed back to the time domain:
\begin{equation}
\tilde{\mathbf{H}}^{l}_{local,t} = \mathcal{F}^{-1}(\tilde{\mathbf{B}}^{l}_{t}) \in \mathbb{R}^{n \times d}.
\end{equation}
We then apply dropout, residual connection, and layer normalization to obtain the $t$-th frequency band output:
\begin{equation} \label{eq:local_output}
\mathbf{o}_{t} = \text{LayerNorm}\left(\mathbf{H}^{l}_{local} + \text{Dropout}(\tilde{\mathbf{H}}^{l}_{local,t})\right).
\end{equation}

To integrate the outputs from all bands, we adopt a soft gating mechanism~\cite{jacobs1991adaptive} that dynamically weighs each band's contribution based on the input representation in the frequency domain. Specifically, a gating function \(g(\cdot)\) is used to produce a softmax distribution over all frequency bands:
\begin{equation} \label{eq:prob_band}
    p_t = \text{Softmax}(g(\|\mathbf{X}^{l}_{local}\|))_t,
\end{equation}
where $\|\mathbf{X}^{l}_{local}\| \in \mathbb{R}^{m\times d}$ denotes the amplitude of the complex spectrum, and $p_t \in \mathbb{R}$ is the importance weight assigned to the $t$-th frequency band. For $g(\cdot)$, we use a three-layer MLP with GeLU activation function.

Unlike sparse gating methods such as top-$k$ selection~\cite{shazeer2017outrageously}, we utilize a soft selection strategy to preserve the contributions of all bands. This avoids potential information loss from discarding seemingly less important frequencies that may still carry valuable user-specific signals.

%Note that we avoid sparse gating~\cite{shazeer2017outrageously} (e.g., selecting only top-$k$ frequency bands) as it would cause information loss by completely discarding potentially relevant frequency bands for certain users.

Finally, the output of LFM is constructed through a weighted aggregation of frequency band outputs based on their learned significance:
\begin{equation}
    \mathbf{O}_{local}^{l} = \sum_{t=1}^K p_t \cdot \mathbf{o}_{t}.
\end{equation}

% 흠..!
% While GFM processes frequencies across whole bands, enabling the capture of diverse user preferences, LFM integrates outputs from different frequency bands, not only effectively capturing important frequency ranges for each user but also providing interpretability for the model's predictions. These two modules serve complementary roles, allowing MUFFIN to effectively handle various behavior patterns ranging from local to global preferences.
% 흠?..!
% This band-specific processing in LFM complements GFM's global spectrum analysis. While GFM captures comprehensive behavioral patterns across the entire frequency spectrum, LFM provides focused modeling of band-specific characteristics, together enabling MUFFIN to overcome both coverage and personalization limitations.

\vspace{1mm}
\noindent
\textbf{Feed forward network (FFN)}. \label{feedfoward}
It incorporates non-linearity into the model, enabling it to capture intricate patterns in user sequences. 
% Specifically:
\begin{align}
    \text{FFN}(\mathbf{X}) = \text{GeLU}(\mathbf{X}\mathbf{W}_1+\mathbf{b}_1)\mathbf{W}_2 + \mathbf{b}_2,
\end{align}
where \(\mathbf{W}_1 \in \mathbb{R}^{d \times 4d}\), \(\mathbf{W}_2 \in \mathbb{R}^{4d \times d}\) are weight matrices, and \(\mathbf{b}_1 \in \mathbb{R}^{4d}, \mathbf{b}_2 \in \mathbb{R}^{d}\) are bias vectors. 

We then apply the FFN to the outputs of both the global and local modules:
\begin{equation}
    \mathbf{H}_{*}^{l+1} = \text{LayerNorm}\left(\mathbf{O}_{*}^{l} + \text{Dropout}\left(\text{FFN}(\mathbf{O}_{*}^{l})\right)\right), 
\end{equation}
\noindent
where $* \in \{ global, \ local \}$.

The updated representations, $\mathbf{H}_{global}^{l+1}$ and $\mathbf{H}_{local}^{l+1}$, are passed into the next layer's GFM and LFM, respectively, allowing the model to progressively refine the sequence representations across layers. After passing the final layer $L$, we obtain the final outputs \(\mathbf{H}^{L}_{global}\) and \(\mathbf{H}^{L}_{local}\), respectively.

\vspace{1mm}
\noindent
\textbf{Prediction layer.} This layer fuses the outputs of both global and local branches to form the final sequence representation used for next-item prediction. We concatenate the final global and local representations and project them through a linear transformation:
\begin{equation}
    \mathbf{H}^{L} = [\mathbf{H}^{L}_{global}; \mathbf{H}^{L}_{local}]\mathbf{W}_p,
\end{equation}
where $\mathbf{W}_p \in \mathbb{R}^{2d \times d}$ and $\mathbf{H}^L \in \mathbb{R}^{n \times d}$.

To preserve information from the original sequence embeddings, we add a residual connection with the initial embedding $\mathbf{H}^{0}$, followed by dropout and layer normalization:
\begin{equation}
    \mathbf{\hat{H}}^{L} = \text{Dropout}\left(\text{LayerNorm}(\mathbf{H}^{0} + \mathbf{H}^{L})\right).
\end{equation}

Finally, we select the representation of the last interacted item \(\mathbf{\hat{H}}^{L}_{|S^{u}|}\in\mathbb{R}^{d}\) and compute the predicted scores for all items.
\begin{align}
    \hat{\mathbf{y}}=\text{Softmax}(\mathbf{E} \mathbf{\hat{H}}^{L}_{|S^{u}|}) \in \mathbb{R}^{|\mathcal{I}|},
\end{align}

\noindent
where $\mathbf{E} \in \mathbb{R}^{|\mathcal{I}|\times d}$ is the whole item embedding matrix shared with the input embedding matrix.

% where \(|I|\) is the total number of items. 

\subsection{Model Training}\label{sec:training}
We train MUFFIN using a multi-objective loss that encourages accurate next-item prediction while promoting the complementary strengths of its dual-module architecture.

\vspace{1mm}
\noindent
\textbf{Recommendation loss.}
The main training objective is to minimize the cross-entropy loss between the predicted distribution \(\hat{\mathbf{y}}\) and the ground truth one-hot vector \(\mathbf{y}\):
\begin{equation}
    \mathcal{L}_{rec} = -\sum_{i=1}^{|\mathcal{I}|} {\mathbf{y}_{i}} \log(\hat{\mathbf{y}}_i).
\end{equation}

\vspace{1mm}
\noindent
\textbf{Auxiliary training loss.} Although the dual-branch design enables MUFFIN to capture both global and local frequency signals, relying solely on the final fused output may not fully exploit the specialization of each module. Motivated by prior work~\cite{wang2023mixture}, we introduce auxiliary supervision that independently encourages each module to perform next-item prediction.

We compute predictions from the final layer representations of each module:
\begin{align}
    \hat{\mathbf{y}}^{local} & = \text{Softmax}(\mathbf{E}  \mathbf{H}_{local}^{L} ), \\
    \hat{\mathbf{y}}^{global} & = \text{Softmax}(\mathbf{E} \mathbf{H}_{global}^{L} ).
\end{align}

The auxiliary loss is defined as the sum of their predictions. This loss encourages global and local modules to develop distinct yet predictive representations.
\begin{align}    
    \mathcal{L}_{aux} = -\sum_{i=1}^{|\mathcal{I}|} \left({\mathbf{y}}_i \log(\hat{\mathbf{y}}_i^{local})+ {\mathbf{y}}_i \log(\hat{\mathbf{y}}_i^{global})\right).
\end{align}

\vspace{1mm}
\noindent
\textbf{Load balancing loss.} Inspired by a study~\cite{zhang2025beyond}, we introduce a load-balancing regularization loss to prevent the soft gate in LFM from focusing disproportionately on a small subset of frequency bands. This loss encourages uniform attention across all $K$ frequency bands by penalizing deviation from the uniform distribution: 
\begin{align}
    \mathcal{L}_{bal} = \frac{1}{K} \sum_{t=1}^K \| p_t-\frac{1}{K} \|^2_2,
\end{align}
where $p_t$ is the soft gate probability of the $t$-th band in Eq.~\eqref{eq:prob_band}. This promotes diversity and full utilization of the frequency spectrum during training.

\vspace{1mm}
\noindent
\textbf{Total loss.}
The overall training objective of MUFFIN combines three loss functions:
\begin{align}
    \mathcal{L} = \mathcal{L}_{rec} + \alpha \mathcal{L}_{aux} + \beta \mathcal{L}_{bal},
\end{align}
where $\alpha$ and $\beta$ are hyperparameters that control the strength of the auxiliary and load balancing losses, respectively.

\begin{table}[] % \small
\caption{Data statistics after preprocessing. `Avg. Length' indicates the average number of interactions per user.}  \label{tab:dataset}
\vspace{-3mm}
\centering
\renewcommand{\arraystretch}{1.0} 
% \footnotesize  % 폰트 크기를 줄여서 표가 넘치는 것을 방지
\begin{tabular}{c|ccccc}
\toprule
Dataset         & Beauty  & Toys  & Sports   & Yelp    & ML-1M  \\ \midrule
\# Users        & 22,363  & 19,412 & 35,598 & 30,499      & 6,041 \\
\# Items        & 12,101  & 11,924  & 18,357 & 20,068      & 3,417 \\
\# Inter.       & 198,502 & 167,597 & 296,337 & 317,182    & 999,611 \\
Avg. Length     & 8.9     & 8.6  & 8.3   & 10.4         & 165.5 \\ 
Sparsity        & 99.93\% & 99.93\% & 99.95\% & 99.95\%   & 95.16\%  \\
\bottomrule
\end{tabular}
\vspace{-2mm}
\end{table}
\section{Experiment Setup}\label{sec:experiments_setup}

\begin{table*}[t]  %\large %\footnotesize
\caption{Accuracy comparison for eight SR models on five datasets. The best and second-best results are marked in \textbf{bold} and \underline{underlined}. $\dagger$ is $p < 0.05$ in a one-tailed t-test, and `Imp.' is improvement ratio, both compared with the second-best model.}
\vspace{-2mm}
\label{tab:overall_perform}
\begin{center}
\renewcommand{\arraystretch}{0.9} % 테이블 행 간격
\small
\begin{tabular}{c|c|cccc|cccc|c|c}
\toprule
Dataset & Metric & GRU4Rec & SASRec & BERT4Rec & DuoRec & FMLPRec & FEARec & BSARec & SLIME4Rec & \textbf{MUFFIN} & Imp. (\%) \\ 
\midrule
\multirow{6}{*}{Beauty} 
& R@5 & 0.0399 & 0.0563 & 0.0338 & 0.0570 & 0.0544 &  {\ul 0.0580} & 0.0557 & 0.0579 &  \textbf{0.0592}$^\dagger$ & \textbf{2.19} \\
& R@10 & 0.0609 & 0.0843 & 0.0484 & 0.0863 & 0.0861 & 0.0863 & 0.0874 &  {\ul 0.0885} &  \textbf{0.0919}$^\dagger$ & \textbf{3.80} \\
& R@20 & 0.0886 & 0.1186 & 0.0696 & 0.1216 & 0.1247 & 0.1227 & 0.1256 &  {\ul 0.1273} &  \textbf{0.1304}$^\dagger$ & \textbf{2.44} \\
& N@5 & 0.0266 & 0.0329 & 0.0232 & 0.0352 & 0.0321 & 0.0360 & 0.0334 &  {\ul 0.0361} &  \textbf{0.0381}$^\dagger$ & \textbf{5.45} \\
& N@10 & 0.0333 & 0.0418 & 0.0280 & 0.0446 & 0.0424 & 0.0451 & 0.0437 &  {\ul 0.0460} &  \textbf{0.0487}$^\dagger$ & \textbf{5.80} \\
& N@20 & 0.0403 & 0.0505 & 0.0333 & 0.0535 & 0.0521 & 0.0545 & 0.0533 &  {\ul 0.0558} &  \textbf{0.0583}$^\dagger$ & \textbf{4.60} \\
 \midrule
\multirow{6}{*}{Toys} 
& R@5 & 0.0339 & 0.0625 & 0.0327 & 0.0653 & 0.0597 & 0.0670 & 0.0636 &  {\ul 0.0670} &  \textbf{0.0692}$^\dagger$ & \textbf{3.28} \\
& R@10 & 0.0510 & 0.0898 & 0.0451 & 0.0946 & 0.0901 & 0.0965 & 0.0944 &  {\ul 0.0995} &  \textbf{0.1005}$^{~~}$ & \textbf{0.97} \\
& R@20 & 0.0740 & 0.1226 & 0.0620 & 0.1296 & 0.1269 & 0.1307 & 0.1310 &  {\ul 0.1365} &  \textbf{0.1382}$^{~~}$ & \textbf{1.22} \\
& N@5 & 0.0235 & 0.0352 & 0.0231 & 0.0388 & 0.0343 & 0.0395 & 0.0379 &  {\ul 0.0401} &  \textbf{0.0441}$^\dagger$ & \textbf{10.06} \\
& N@10 & 0.0290 & 0.0440 & 0.0271 & 0.0483 & 0.0441 & 0.0490 & 0.0478 &  {\ul 0.0506} &  \textbf{0.0542}$^\dagger$ & \textbf{7.25} \\
& N@20 & 0.0347 & 0.0522 & 0.0313 & 0.0571 & 0.0534 & 0.0576 & 0.0571 &  {\ul 0.0599} &  \textbf{0.0641}$^\dagger$ & \textbf{7.02} \\
 \midrule
\multirow{6}{*}{Sports} 
& R@5 & 0.0225 & 0.0303 & 0.0118 & 0.0322 & 0.0322 & 0.0286 & 0.0320 &  {\ul 0.0341} &  \textbf{0.0359}$^\dagger$ & \textbf{5.18} \\
& R@10 & 0.0362 & 0.0476 & 0.0206 & 0.0498 & 0.0511 & 0.0434 & 0.0511 &  {\ul 0.0530} &  \textbf{0.0570}$^\dagger$ & \textbf{7.48} \\
& R@20 & 0.0566 & 0.0690 & 0.0345 & 0.0734 & 0.0759 & 0.0635 & 0.0751 &  {\ul 0.0788} &  \textbf{0.0841}$^\dagger$ & \textbf{6.64} \\
& N@5 & 0.0150 & 0.0167 & 0.0073 & 0.0198 & 0.0178 & 0.0176 & 0.0177 &  {\ul 0.0210} &  \textbf{0.0213}$^{~~}$ & \textbf{1.43} \\
& N@10 & 0.0193 & 0.0223 & 0.0101 & 0.0254 & 0.0239 & 0.0223 & 0.0239 &  {\ul 0.0271} &  \textbf{0.0281}$^\dagger$ & \textbf{3.82} \\
& N@20 & 0.0244 & 0.0276 & 0.0136 & 0.0313 & 0.0301 & 0.0274 & 0.0299 &  {\ul 0.0336} &  \textbf{0.0349}$^\dagger$ & \textbf{3.87} \\
 \midrule
\multirow{6}{*}{Yelp} 
& R@5 & 0.0253 & 0.0425 & 0.0192 & 0.0435 & 0.0471 & 0.0399 & 0.0458 &  {\ul 0.0473} &  \textbf{0.0510}$^\dagger$ & \textbf{7.75} \\
& R@10 & 0.0415 & 0.0597 & 0.0341 & 0.0634 & 0.0681 & 0.0567 & 0.0691 &  {\ul 0.0729} &  \textbf{0.0755}$^{~~}$ & \textbf{3.66} \\
& R@20 & 0.0671 & 0.0862 & 0.0570 & 0.0929 & 0.0998 & 0.0817 & 0.1030 &  {\ul 0.1097} &  \textbf{0.1111}$^{~~}$ & \textbf{1.31} \\
& N@5 & 0.0166 & 0.0322 & 0.0120 & 0.0317 & {\ul 0.0344} & 0.0287 &  0.0325 & 0.0323 &  \textbf{0.0350}$^\dagger$ & \textbf{1.65} \\
& N@10 & 0.0217 & 0.0377 & 0.0167 & 0.0381 & {\ul 0.0412} & 0.0341 & 0.0400 &  0.0405 &  \textbf{0.0429}$^\dagger$ & \textbf{4.21} \\
& N@20 & 0.0282 & 0.0444 & 0.0225 & 0.0455 & 0.0491 & 0.0404 & 0.0485 &  {\ul 0.0498} &  \textbf{0.0518}$^\dagger$ & \textbf{3.83} \\
 \midrule
\multirow{6}{*}{ML-1M} 
& R@5 & 0.2085 &  {\ul 0.2182} & 0.1846 & 0.2168 & 0.2160 & 0.1843 & 0.2168 & 0.2158 &  \textbf{0.2252}$^\dagger$ & \textbf{3.22} \\
& R@10 & 0.2925 &  {\ul 0.3103} & 0.2722 & 0.3098 & 0.3049 & 0.2709 & 0.3090 & 0.3044 &  \textbf{0.3187}$^\dagger$ & \textbf{2.71} \\
& R@20 & 0.3924 &  {\ul 0.4121} & 0.3823 & 0.4099 & 0.4078 & 0.3829 & 0.4117 & 0.4080 &  \textbf{0.4179}$^\dagger$ & \textbf{1.42} \\
& N@5 & 0.1436 & 0.1495 & 0.1222 &  {\ul 0.1497} & 0.1487 & 0.1237 & 0.1485 & 0.1479 &  \textbf{0.1546}$^\dagger$ & \textbf{3.27} \\
& N@10 & 0.1707 & 0.1791 & 0.1505 &  {\ul 0.1798} & 0.1774 & 0.1517 & 0.1782 & 0.1765 &  \textbf{0.1848}$^\dagger$ & \textbf{2.78} \\
& N@20 & 0.1959 & 0.2049 & 0.1783 &  {\ul 0.2050} & 0.2034 & 0.1799 & 0.2042 & 0.2026 &  \textbf{0.2098}$^\dagger$ & \textbf{2.36} \\
\bottomrule
\end{tabular}
\vspace{-3mm}
\end{center}
\end{table*}

% \subsection{Datasets}
\textbf{Datasets}.
We evaluate MUFFIN on five widely used benchmark datasets for sequential recommendation (SR), following the standard setup in prior work~\cite{DuYZF0LS023SLIME4Rec}. These datasets span diverse domains -- e-commerce, local services, and media -- providing a comprehensive validation of the model's effectiveness.
\textbf{Amazon Beauty, Toys, and Sports}\footnote{\url{https://cseweb.ucsd.edu/~jmcauley/datasets/amazon/links.html}}~\cite{sigir/McAuleyTSH15Amazonreview} are subsets of the Amazon product review corpus. Each dataset contains timestamped user-item interactions derived from product reviews and ratings.
\textbf{Yelp}\footnote{\url{https://www.yelp.com/dataset}} consists of user interactions with local businesses, including reviews and ratings.
\textbf{ML-1M}\footnote{\url{https://grouplens.org/datasets/movielens/}} contains 1 million movie ratings from 6,000 users on 4,000 movies, along with user demographics and movie metadata information.

% \begin{itemize}[leftmargin=5mm]
%     \item\textbf{Amazon Beauty, Toys, and Sports}\footnote{\url{https://cseweb.ucsd.edu/~jmcauley/datasets/amazon/links.html}}~\cite{sigir/McAuleyTSH15Amazonreview}: These datasets are subsets of the Amazon product review corpus. Each dataset contains timestamped user-item interactions derived from product reviews and ratings, making them suitable for studying user behavior in e-commerce settings.
    
%     \item\textbf{Yelp}\footnote{\url{https://www.yelp.com/dataset}}: This dataset consists of user interactions with local businesses, including reviews and ratings. It reflects real-world sequential user behavior in the context of local services and is commonly used in SR tasks.

%     \item\textbf{ML-1M}\footnote{\url{https://grouplens.org/datasets/movielens/}}: This is a widely used benchmark for recommendation systems. Due to its clean structure and popularity, it serves as a standard benchmark for evaluating sequential recommendation models in the entertainment domain.
% \end{itemize}

Each dataset is preprocessed following established protocols to ensure fair and consistent comparisons with prior work~\cite{SunLWPLOJ19BERT4Rec, DuYZQZ0LS23FEARec, DuYZF0LS023SLIME4Rec}. Specifically, we use \emph{five-core settings} by removing users and items that occur less than five times in the dataset. Table~\ref{tab:dataset} reports the detailed statistics of datasets.

\vspace{1mm}
\noindent
\textbf{Competing models}. We compare MUFFIN against eight SR baselines, including time-domain and frequency-domain SR models.

\noindent
\textbf{\emph{Time-domain SR models: }}
\textbf{GRU4Rec}~\cite{HidasiKBT15GRU4Rec} utilizes a gated recurrent unit to model users' sequential behaviors.
\textbf{SASRec}~\cite{KangM18SASRec} leverages the self-attention mechanism to model item-item dependencies within sequences.
\textbf{BERT4Rec}~\cite{SunLWPLOJ19BERT4Rec} applies BERT's bidirectional encoding structure to SR, improving the contextual understanding of sequences through masked item prediction.
\textbf{DuoRec}~\cite{QiuHYW22DuoRec} enhances sequence representations by incorporating contrastive self-supervised learning tasks into the transformer architecture.

\noindent
\textbf{\emph{Frequency-domain SR models: }}
\textbf{FMLP-Rec}~\cite{ZhouYZW22FMLPRec} is a pioneer frequency-domain SR model using element-wise complex weight filters to process frequency components.
\textbf{FEARec}~\cite{DuYZQZ0LS23FEARec} integrates frequency-domain signals directly into attention computation, enabling the model to leverage time and frequency information.

\noindent
\textbf{BSARec}~\cite{aaai/Shin0WP24BSARec} adjusts the influence on the high-frequency domains and combines it with an inductive bias of the self-attention mechanism.
\textbf{SLIME4Rec}~\cite{DuYZF0LS023SLIME4Rec} utilizes a frequency ramp structure with layer-wise dynamic and static frequency band selection to capture diverse sequential patterns.

\vspace{1mm}
\noindent
\textbf{Evaluation protocol}.
We evaluate the performance of the next-item prediction task using two standard metrics:  Recall@K and NDCG@K, where K is set to \{5, 10, 20\}. For brevity, we denote these metrics as R@K and N@K, respectively. Following prior works~\cite{rendle2012bpr, KangM18SASRec}, we adopt the \emph{leave-one-out evaluation} strategy. For each user’s interaction sequence, the last item is held out as the test instance, the second-to-last item is used for validation, and the remaining interactions form the training set. We rank the ground-truth item against all other items (including those that appeared in the training set) and compute the ranking-based metrics on the test set. The final results are the average scores across all test users. All the results are averaged over five runs with different random seeds.

% \subsection{Implementation Details}
\vspace{1mm}
\noindent
\textbf{Implementation details}.
All experiments were conducted using the Recbole\footnote{\url{https://github.com/RUCAIBox/RecBole}}~\cite{recbole} and Recbole-DA\footnote{\url{https://github.com/RUCAIBox/RecBole-DA}} frameworks, open-source libraries for building and evaluating SR models. MUFFIN was implemented on Recbole-DA to ensure a fair comparison, as it is the same experimental environment used in SLIME4Rec~\cite{DuYZF0LS023SLIME4Rec}. For all models, we adopt the Adam optimizer~\cite{KingmaB14Adam} with a learning rate of 0.001, a batch size of 256, and a hidden dimension of 64. To ensure consistency and comparability with prior works~\cite{DuYZF0LS023SLIME4Rec,KangM18SASRec,aaai/Shin0WP24BSARec,DuYZQZ0LS23FEARec,ZhouYZW22FMLPRec}, all frequency-based models are set to use two layers. To train MUFFIN, we conduct a grid search for hyperparameters: the number of frequency bands $K \in \{2, 4, 6, 8, 10\}$, the auxiliary loss weight $\alpha \in \{0.05, 0.1, 0.2, 0.5, 1\}$, and load balancing loss weight $\beta \in \{0.05, 0.1, 0.2, 0.5, 1\}$. The dropout rate is set to 0.1 for ML-1M and 0.4 for the remaining datasets. We searched for the kernel size $c$ for the UAF module over \{3, 5, 7\}. All baseline models were trained with the hyperparameters reported in their original papers to ensure optimal performance. We adopt the cross-entropy loss function across all models, as it consistently outperforms BCE and BPR loss in prior studies. Following standard practice~\cite{DuYZQZ0LS23FEARec, aaai/Shin0WP24BSARec}, the maximum sequence length is set to 50. N@20 is used as the validation metric, and early stopping is applied if performance does not improve for 15 consecutive epochs. All experiments were conducted on a server equipped with an NVIDIA RTX 3090 24GB GPU and an Intel Xeon Gold 6226R CPU.

% , including SLIME4Rec~\cite{DuYZF0LS023SLIME4Rec}\footnote{Note that, because it was initially implemented with eight layers, our reported result is different from the original work~\cite{DuYZF0LS023SLIME4Rec}.}

% \vspace{0.5mm}
% All experiments were performed using Recbole~\cite{} framework, which is an open-source library for recommendation systems. We adopt Adam optimizer ~\cite{} with a learning rate of 0.001. Batch size 256 hidden size 64. We set the maximum sequence length to 50, following the conventions~\cite{}. Validation metric으로는 NDCG@20을 사용하여 15회 동안 성능이 오르지 않는다면 학습을 중단하도록 하였다. gamma는 [0.01, 0.05, 0.1, 0.2, 0.5], aux\_weight은 [0.05, 0.1, 0.2, 0.5, 1], dropout ratio는 [0.1, 0.2, 0.3, 0.4, 0.5]로 진행하였다. 다른 Baseline model들은 해당 paper에서 report한 ideal 성능이 나오도록 hyper-parameter를 참고하여 실험을 진행하였다. 

\section{Experimental Results}

\subsection{Overall Performance}
Table~\ref{tab:overall_perform} reports the performance of MUFFIN and other baseline models across five datasets. All improvement ratios are computed from exact values without rounding. Our key findings are as follows. 

\vspace{1mm}
\noindent
\textbf{Outstanding performance of MUFFIN}. MUFFIN consistently achieves state-of-the-art performance across all datasets. Compared to DuoRec~\cite{QiuHYW22DuoRec} as the strongest transformer-based baseline, MUFFIN achieves average relative improvements of 9.84\% in R@10 and 9.48\% in N@10. Against SLIME4Rec~\cite{DuYZF0LS023SLIME4Rec} as the strongest frequency-domain competitor, MUFFIN achieves average gains of 4.12\% in R@10 and 5.48\% in N@10. These improvements underscore MUFFIN’s effectiveness in overcoming the limitations of prior models via its dual filtering architecture, simultaneously capturing both global and local behavioral patterns.

\vspace{1mm}
\noindent
\textbf{Time-domain vs. Frequency-domain models.} Traditional time-domain models like, SASRec~\cite{KangM18SASRec} and DuoRec~\cite{QiuHYW22DuoRec} achieve competitive results. Their performances are particularly pronounced on ML-1M with long user sequences. In this situation, their superior capability to model long-range dependencies proves advantageous. Recent frequency-domain models show promising advancements over traditional time-domain models. Among them, SLIME4Rec~\cite{DuYZF0LS023SLIME4Rec} stands out, outperforming most time-domain models on four out of five datasets. This highlights the potential of frequency-aware modeling in capturing intricate user behavior patterns.

% Table ~\ref{tab:overall_perform}는 MUFFIN과 다른 베이스라인 모델들의 성능을 5개의 dataset에 대해 나타낸다.  모든 모델의 성능은 5개의 random seed에서의 평균을 나타낸다. 가장 우수한 성능은 bold를, 두번째로 우수한 모델은 underline을 하였다. 우리가 발견한 몇 가지 키 포인트들은 다음과 같다. (i) Transformer기반 모델은 우수한 성능을 보인다. 우리는 다른 논문과 달리 모든 모델의 loss function은 BPR대신 Cross Entropy loss를 사용하였다. 이로인해 SASRec의 경우, 최근 state-of-the-art모델과 경쟁가능한 수준의 성능을 보이고 있다. (ii) 그럼에도 불구하고, Self attention기반 모델보다 우수한 성능을 보여주는 Frequency based model들이 여럿 등장하였다. BSARec의 경우 SASRec 대비 ML-1m을 제외한 데이터셋에서 R@10에서 평균적으로 6.29\%의 gain, N@10에서 평균적으로 5.20\%의 gain을 보여준다. (iii) Contrastive learning을 활용하는 모델 또한 높은 성능을 보인다. DuoRec은 SASRec 대비 R@10에서 평균적으로 3.66\% gain, N@10에서 6.38\% gain을 보여주었고 SLIME4Rec은 BSARec대비 R@10에서 평균적으로 2.90\% gain을  N@10에서 4.93\%의 gain을 보여주었다.  (iv) MUFFIN은 지속적으로 모든 dataset에서 state-of-the-art 성능을 보여준다. transformer 기반 모델 중 가장 뛰어난 DuoRec과 비교했을 때 R@10과 N@10에서 각각 평균 9.84\%, 9.41\%의 성능 향상을 보여주었다. 또한 Frequency Domain에서 학습한 가장 경쟁적인 모델인 SLIME4Rec과 비교해 봤을 때도 평균적으로 R@10에서 4.06\%, N@10에서 5.37\%의 성능향상을 보여주었다. 특히 Yelp에서는 R@5와 N@5에서 10\%의 성능 향상을 보여주었다. 이러한 실험 결과는 Muffin이 

\subsection{Ablation Study}
To evaluate the effect of each component in MUFFIN, we conduct extensive ablation studies on the Beauty and ML-1M datasets. Table~\ref{tab:ablation} reports the comparative results of the original model and its ablated variants.

\vspace{1mm}
\noindent
\textbf{Effect of key model components.}
Removing any core component in MUFFIN resulted in consistent performance degradation, highlighting the importance of each architectural element. Notably, substituting user-adaptive filtering (UAF) with fixed non-adaptive filters (\emph{w/o UAF}) leads to performance drops, underscoring the value of personalized frequency selection. Additionally, eliminating either the Global Filtering Module (GFM) or the Local Filtering Module (LFM) (\emph{w/o GFM} or \emph{w/o LFM}) significantly reduces performance compared to the complete dual-module configuration. Based on these results, we confirm that global and local frequency perspectives are complementary and jointly necessary for modeling complex user behavior.

\vspace{1mm}
\noindent
\textbf{Effect of loss functions.} The auxiliary training loss is also critical to the model's effectiveness. Removing the auxiliary loss $\mathcal{L}{aux}$, which provides independent supervision to GFM and LFM, results in the most substantial drop in performance. This highlights its importance in promoting the distinct learning objectives of the two modules. Similarly, removing the load balancing loss $\mathcal{L}{bal}$ degrades performance, indicating its role in ensuring effective and balanced utilization of frequency bands. Further analysis of the load balancing mechanism is provided in Section~\ref{sec:indepth}.

\vspace{1mm}
\noindent\textbf{Effect of convolutional filters in UAF.} To validate the effectiveness of the convolutional filter design in UAF, we conduct an additional study by replacing the convolution layers with MLP layers (denoted as $\text{MUFFIN}_{\text{MLP}}$ ). As reported in Table~\ref{tab:UAF2MLP}, MUFFIN using convolutional filters consistently outperforms $\text{MUFFIN}_{\text{MLP}}$ on both datasets. Unlike MLPs that process each frequency component independently, convolutional filters better capture local frequencies and spatial patterns in the frequency spectrum. This capability makes MUFFIN more effective for generating user-specific filters.

%Both auxiliary loss components contribute significantly to MUFFIN's performance. The auxiliary loss ($\mathcal{L}_{aux}$) showed the most critical impact, with its removal causing substantial degradation in R@10 and demonstrating that the independent supervision of both GFM and LFM modules is crucial for effective representation learning. The load balancing loss ($\mathcal{L}_{bal}$) also proved essential, as its absence reduced the model's ability to leverage diverse frequency bands effectively. Further analysis of the load balancing mechanism is provided in Section~\ref{sec:frequency band prob}.

% All variant models with removed core components (UAF, GFM, LFM) showed performance degradation compared to the original MUFFIN. We observed performance drops when replacing user-adaptive filtering with a fixed filter (w/o UAF) and using a single module (w/o GFM, w/o LFM). Notably, the performance degradation when using a single module demonstrates the effectiveness of MUFFIN's approach in mixing dual filtering modules for enhanced recommendation performance.

% The results comprehensively demonstrate that each component of MUFFIN makes unique contributions to the overall model performance.

\begin{table}[]  %\large %\footnotesize
\caption{Ablation study of MUFFIN on Beauty and ML-1M. `w/o' denotes the model variant without the corresponding component.}
\vspace{-2mm}
\label{tab:ablation}
\renewcommand{\arraystretch}{1} % 테이블 행 간격

\begin{tabular}{c|cc|cc}
\toprule
\multicolumn{1}{c|}{\multirow{2}{*}{Model}} & \multicolumn{2}{c|}{Beauty} & \multicolumn{2}{c}{ML-1M} \\
 & R@10 & N@10 & R@10 & N@10 \\
\midrule
MUFFIN & 0.0919 & 0.0487 & 0.3187 & 0.1848 \\
\midrule
\ \ \ \ \ \ \ \ w/o UAF \ \ \ & 0.0906 & 0.0473 & 0.3151  & 0.1803 \\
\ \ \ \ \ \ \ \ w/o GFM \ \ \ & 0.0887 & 0.0460 & 0.3086  & 0.1759  \\
\ \ \ \ \ \ \ \ w/o LFM \ \ \ & 0.0906 & 0.0467  & 0.3079 & 0.1798  \\
\midrule
\ \ \ \ \ \ \ \ w/o $\mathcal{L}_{aux}$ \ \ \ & 0.0862 & 0.0471 & 0.3101 & 0.1813 \\
\ \ \ \ \ \ \ \ w/o $\mathcal{L}_{bal}$ \ \ \ & 0.0907 & 0.0475 & 0.3151 & 0.1810 \\
\bottomrule
\end{tabular}
\vspace{-2mm}
\end{table}

\begin{table}[]  %\large %\footnotesize
\caption{Accuracy comparison between MUFFIN and $\textbf{MUFFIN}_{\text{MLP}}$ on Beauty and ML-1M. $\textbf{MUFFIN}_{\text{MLP}}$ replaces the convolution layers with MLP layers in UAF.}
\label{tab:UAF2MLP}
\vspace{-2mm}
\renewcommand{\arraystretch}{1} % 테이블 행 간격
\begin{tabular}{c|cc|cc}
\toprule
\multicolumn{1}{c|}{\multirow{2}{*}{Model}}  & \multicolumn{2}{c|}{Beauty} & \multicolumn{2}{c}{ML-1M} \\
 & R@10 & N@10 & R@10 & N@10 \\
\midrule
MUFFIN (ours) & 0.0919 & 0.0487 & 0.3187 & 0.1848  \\
% \midrule
$\text{MUFFIN}_{\text{MLP}}$  & 0.0913 & 0.0476 & 0.3115 & 0.1810 \\
\bottomrule
\end{tabular}
\vspace{-3mm}
\end{table}

\begin{figure}
%\vspace{2mm}
\includegraphics[height=0.4cm]{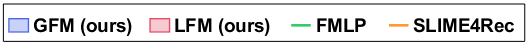}
\vspace{-3mm}
\begin{tabular}{cc}
%\vspace{-1mm}
\includegraphics[height=3.1cm]{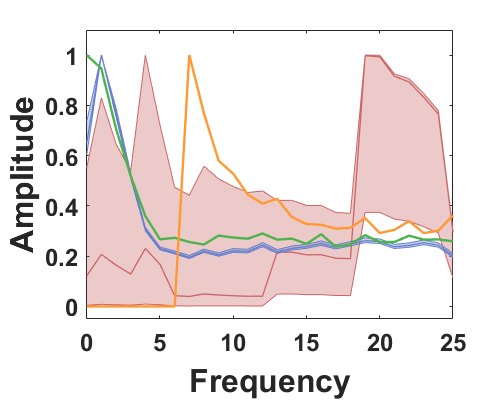} &
\includegraphics[height=3.1cm]{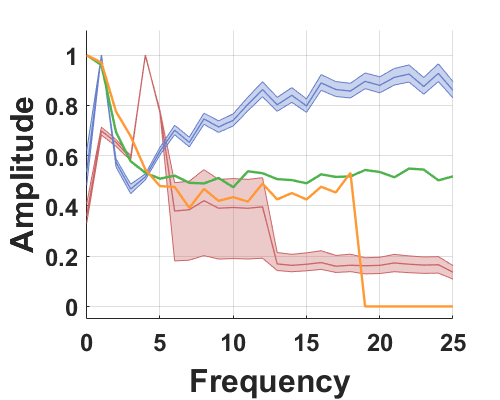} \\
\multicolumn{1}{c}{(a) Layer 1 filter (Beauty)} &\multicolumn{1}{c}{(b) Layer 2 filter (Beauty)} \\
%\vspace{-1mm}
\includegraphics[height=3.1cm]{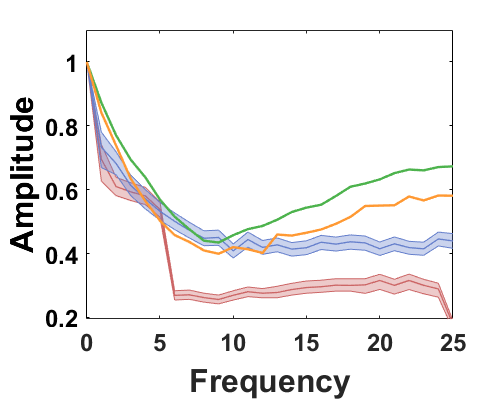} &
\includegraphics[height=3.1cm]{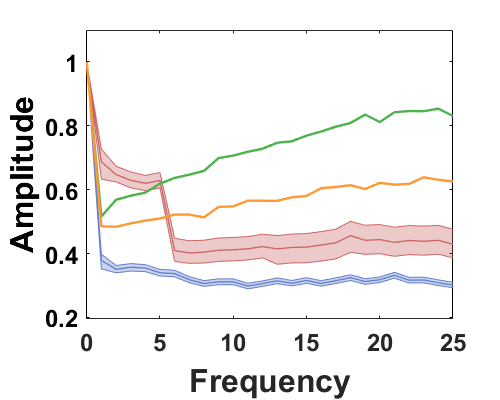} \\
\multicolumn{1}{c}{(c) Layer 1 filter (ML-1M)} &\multicolumn{1}{c}{(d) Layer 2 filter (ML-1M)}
\end{tabular}

\caption{Frequency filter amplitude distributions of MUFFIN's modules and baseline models on Beauty and ML-1M. The x-axis represents frequency indices, and the y-axis represents filter amplitude values. Shaded areas for MUFFIN's modules illustrate the variation in amplitude across different users for each frequency component. }\label{fig:freq_analysis3}
\vspace{-3mm}
\end{figure}
\subsection{In-depth Analysis}\label{sec:indepth}

\textbf{Visualization of different frequency filters.} To investigate how user-adaptive frequency filtering in MUFFIN differs from existing models, we analyze the learned frequency filters of representative models. Figure~\ref{fig:freq_analysis3} shows the user-adaptive filters of GFM and LFM in MUFFIN compared to the frequency filters of FMLPRec~\cite{ZhouYZW22FMLPRec} and SLIME4Rec~\cite{DuYZF0LS023SLIME4Rec}. We found two interesting findings as follows. (i) FMLPRec~\cite{ZhouYZW22FMLPRec} exhibits different tendencies across datasets. It emphasizes low-frequency components (indices 0--5) in Beauty while showing a gradually increasing pattern toward high-frequency regions in ML-1M. SLIME4Rec~\cite{DuYZF0LS023SLIME4Rec} completely blocks certain frequency bands (\ie, zero amplitude) in Beauty while maintaining intermediate values across all frequency bands in ML-1M. Moreover, both FMLPRec~\cite{ZhouYZW22FMLPRec} and SLIME4Rec~\cite{DuYZF0LS023SLIME4Rec} employ identical filters for all users. (ii) In clear contrast, both GFM and LFM in MUFFIN exhibit variability in filter weights even at the same frequency components, indicating dynamically generated user-adaptive filters. Specifically, GFM maintains relatively stable patterns over the full frequency spectrum while incorporating user-specific variations. Meanwhile, on Beauty, LFM assigns highly diverse weights ranging from near zero to one across low and high-frequency bands among users. On ML-1M, LFM shows user-specific variations predominantly in mid to high-frequency bands. These findings highlight that MUFFIN mitigates the static filtering of existing frequency-domain models by dynamically generating user-adaptive frequency filters that reflect each user’s unique frequency characteristics.

% MUFFIN의 사용자 적응적 주파수 필터링이 기존 접근법과 어떻게 다른지 이해하기 위해, 서로 다른 모델과 데이터셋에서 학습된 주파수 필터들을 분석했다. Figure 5는 베이스라인 모델인 FMLPRec과 SLIME4Rec의 고정된 주파수 필터와 MUFFIN의 GFM과 LFM 모듈의 사용자 적응적 주파수 필터를 보여준다. 
% Figure 5에서 볼 수 있듯이 FMLPRec의 경우 데이터셋별로 뚜렷한 선호도 차이를 나타내는데, Beauty 데이터셋에서는 저주파 성분(0-5 인덱스)을 강조하는 반면, ML-1M에서는 고주파 영역으로 갈수록 점진적으로 증가하는 패턴을 보인다. SLIME4Rec은 데이터셋에 따라 서로 다른 극단적 정적 필터링을 수행하는데, Beauty에서는 특정 주파수 대역을 완전히 차단(amplitude 0)하는 반면, ML-1M에서는 전 주파수 대역에 걸쳐 상대적으로 보수적인 중간값을 유지한다. 이러한 기존 방법들은 모든 사용자에게 동일한 필터를 적용한다.
% 이와 대조적으로, MUFFIN의 두 모듈은 다른 특성을 보인다. 기존 방법들이 각 주파수에서 고정된 단일값을 갖는 반면, MUFFIN의 GFM과 LFM은 동일한 주파수 성분에서도 상당한 분산을 가진 분포를 나타낸다. 이러한 분산 특성은 각 사용자별로 개별화된 필터 가중치가 동적으로 생성됨을 의미한다. GFM은 전체 주파수 스펙트럼에 걸쳐 상대적으로 안정적인 패턴을 유지하면서도 사용자별로 변동성을 나타난다. 반면, LFM은 극도의 개인화를 보여주는데, Beauty에서는 저주파와 고주파 영역에서 사용자별로 0에 가까운 값부터 1에 가까운 값까지 극도로 다양한 가중치를 학습하고, ML-1M에서는 중간 주파수 대역에서 고주파수 대역까지 사용자별 변동성을 나타낸다.
% 또한 계층별 두 모듈의 필터가 다른 특성을 가지는 것이 나타난다. Beauty 데이터셋에서 layer 1에서 GFM이 고주파 영역을 강조하는 반면 첫번째 레이어에서는 LFM이 고주파 영역을 강조한다. ML-1M에서는 layer 1에서 GFM이 LFM보다 상대적으로 고주파를 더 강조하며, layer2 에서는 반대되는 경향을 보인다. 이는 두 모듈이 차별화된 주파수 처리를 통해 서로 상호 보완적으로 작동하는 것을 보인다. 이러한 결과는 MUFFIN이 기존 방법들의 정적 필터링 한계를 성공적으로 극복하고, 각 사용자의 고유한 주파수 특성에 따라 동적으로 개인화된 필터를 생성함을 보여준다.

% \input{Figures_tex/frequency_analysis2}

\begin{figure}
\centering
\begin{tabular}{cc}
\includegraphics[width=0.45\linewidth]{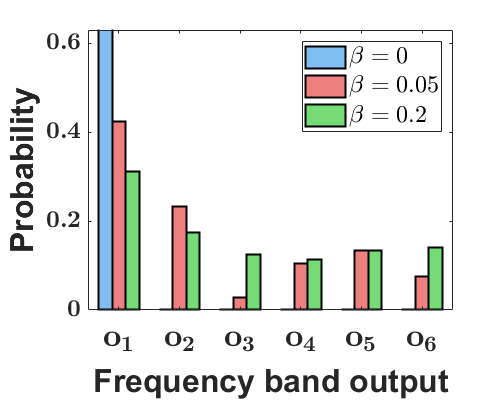} & \includegraphics[width=0.45\linewidth]{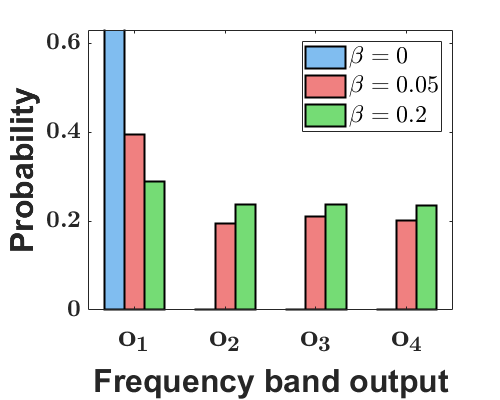} \\
\multicolumn{1}{c}{(a) Beauty} &\multicolumn{1}{c}{(b) ML-1M} \\
\end{tabular}
\vspace{-2mm}
\caption{Probability distribution of frequency band outputs in LFM across different load balancing weights $\beta$ on Beauty and ML-1M.}\label{fig:band_prob}
\vspace{-2mm}
\end{figure}

\begin{table}[]  %\small %\footnotesize
\caption{Efficiency comparison (in seconds) on Beauty and ML-1M. Training time is measured per epoch; evaluation time is measured for all test users.}
\vspace{-2mm}
\label{tab:complexity}
\renewcommand{\arraystretch}{1} % 테이블 행 간격

\begin{tabular}{c|cc|cc}
\toprule
\multicolumn{1}{c|}{\multirow{1}{*}{Dataset}} & \multicolumn{2}{c|}{Beauty} & \multicolumn{2}{c}{ML-1M} \\
\midrule
Model & Train & Eval & Train & Eval \\
\midrule
DuoRec    & 23.3 & 0.18 & 161.2 & 0.17  \\
BSARec    & 13.5 & 0.49 & 85.0 & 0.19  \\
SLIME4Rec & 31.7 & 0.16 & 253.6 & 0.13 \\
MUFFIN (ours)  & 15.4 & 0.67 & 131.5 & 0.21 \\
\bottomrule
\end{tabular}
\vspace{-2mm}
\end{table}

\vspace{1mm}
\noindent\textbf{Local frequency band probability.} We analyze the probability distributions of frequency bands in LFM under varying load-balancing weights $\beta$. Figure~\ref{fig:band_prob} depicts the average probability distribution across LFM frequency bands when trained with different $\beta$ values on Beauty and ML-1M. The x-axis represents frequency band outputs $\mathbf{o}_t$ in Eq.~\eqref{eq:local_output} ordered from the lowest frequency band output to the highest, while the y-axis indicates the corresponding probabilities. We use six and four frequency bands in Beauty and ML-1M, respectively, based on their optimal configurations (see Section~\ref{sec:hyper_param}).
For Beauty, as shown in Figure~\ref{fig:band_prob}(a), the probability distribution is well balanced across bands $\mathbf{o}_2$ to $\mathbf{o}_6$ at $\beta=0.2$ (optimal hyperparameter), demonstrating that MUFFIN effectively mitigates the limited frequency band coverage issue by leveraging a diverse range of frequency components. In contrast, without load balancing loss ($\beta=0$), LFM heavily concentrates on $\mathbf{o}_1$, \ie, specific low-frequency ranges, thereby leading to the loss of crucial behavioral information. For ML-1M,  Figure~\ref{fig:band_prob}(b) shows similar trends. Notably, low-frequency bands maintain consistently high probabilities regardless of $\beta$, reflecting their fundamental role in capturing stable user preferences. Meanwhile, the balanced utilization of high-frequency bands at the optimal $\beta$ facilitates personalized modeling of diverse behavioral variations across both datasets.

\vspace{1mm}
\noindent\textbf{Training and evaluation cost.} We analyze the computational efficiency of MUFFIN compared to competing models. 
As shown in Table~\ref{tab:complexity}, MUFFIN demonstrates superior training efficiency compared to the strongest baseline SLIME4Rec~\cite{DuYZF0LS023SLIME4Rec}, achieving approximately 2$\times$ and 1.6$\times$ faster training time on Beauty and ML-1M, respectively.
While SLIME4Rec incurs substantial training overhead due to complicated augmentation and loss computations for contrastive learning, MUFFIN saves significant training time.
% Meanwhile, although MUFFIN shows higher evaluation costs than some baselines, this overhead is minimal at only 0.03--0.035ms per user (0.03ms for Beauty, 0.035ms for ML-1M). In practical scenarios where recommendations are typically generated for individual users or small batches, this marginal overhead is acceptable compared to the consistent accuracy improvements.
For evaluation time, MUFFIN shows increased costs compared to baselines, with total evaluation times of 0.67s and 0.21s on Beauty and ML-1M, respectively, compared to the fastest baseline (\ie, SLIME4Rec) times of 0.16s and 0.13s. When calculated per user, this is approximately 0.03-0.035ms of additional time, which stems from the computational overhead of the dual filtering architecture. The increase in evaluation time reflects the trade-off between computational efficiency and the accuracy improvements.

% Table~\ref{tab:complexity} shows the training and evaluation times on the Beauty and ML-1M datasets. For training time, it demonstrates remarkable efficiency. MUFFIN achieves approximately 1.6$\times$ speed improvement with 131.5 seconds compared to SLIME4Rec~\cite{DuYZF0LS023SLIME4Rec} (253.6 seconds) on the ML-1M dataset. The Beauty dataset shows an even more pronounced difference with 15.4 seconds, achieving approximately 2$\times$ faster training speed than SLIME4Rec (31.7 seconds).

\subsection{Hyper-parameter Sensitivity} \label{sec:hyper_param}
We evaluate the effect of three key hyperparameters used in MUFFIN on model accuracy. For hyperparameter analysis, experiments are conducted with a single seed.

\begin{figure}
\begin{tabular}{cc}
\includegraphics[height=3.0cm]{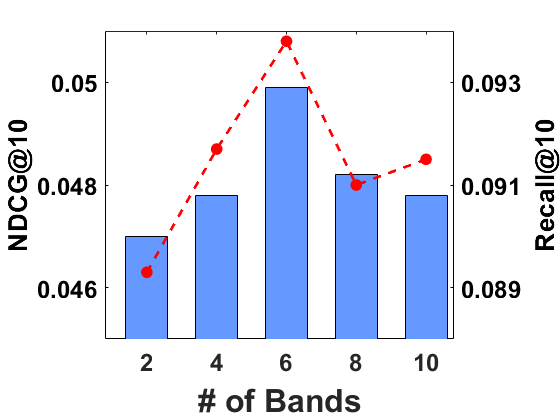} &
\includegraphics[height=3.0cm]{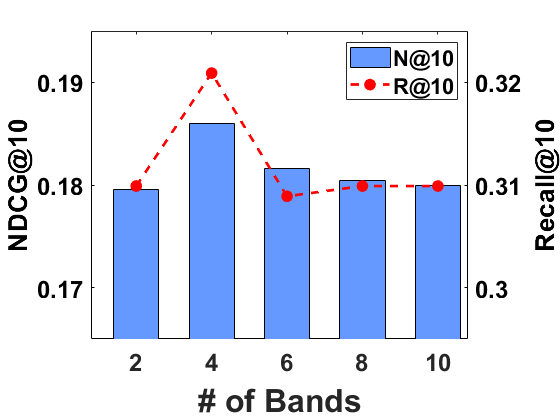} \\
\multicolumn{1}{c}{(a) Beauty} &\multicolumn{1}{c}{(b) ML-1M} \\
\end{tabular}
\vspace{-3mm}
\caption{Accuracy of MUFFIN over varying the number of frequency bands on Beauty and ML-1M. }\label{fig:hyper_experts}
\vspace{-2mm}
\end{figure}

\vspace{1mm}
\noindent
\textbf{Number of frequency bands $K$.}
In Figure~\ref{fig:hyper_experts}, we analyzed performance while varying the number of frequency bands ($K$). When $K=1$, LFM is equivalent to GFM. Performance improved with increasing $K$, peaking at $K=4, 6$. Both too few or too many frequency bands led to suboptimal performance, as insufficient bands limit fine-grained pattern capture while excessive bands may cause the overfitting problem.

\begin{figure}
\begin{tabular}{cc}
\includegraphics[height=3.0cm]{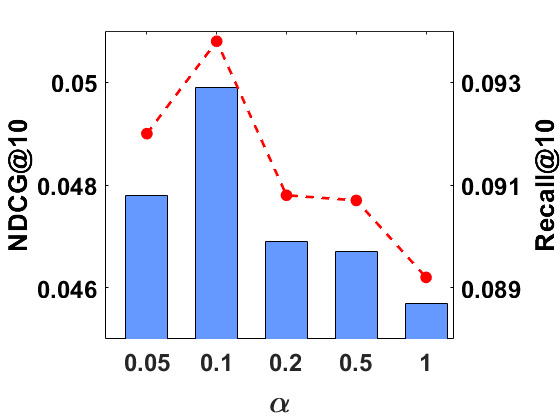} &
\includegraphics[height=3.0cm]{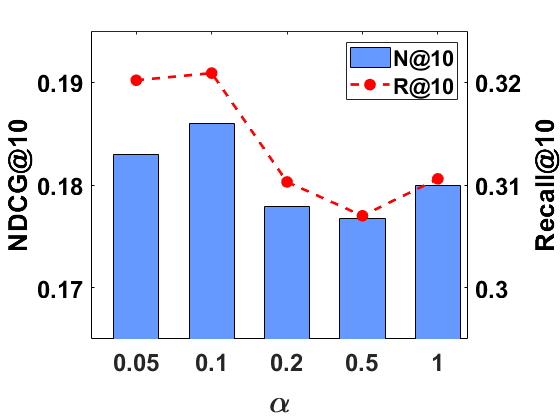} \\
\multicolumn{1}{c}{(a) Beauty} &\multicolumn{1}{c}{(b) ML-1M}
\end{tabular}
\vspace{-3mm}
\caption{Accuracy of MUFFIN over varying auxiliary loss weight $\alpha$ on Beauty and ML-1M. }\label{fig:hyper_aux}
\vspace{-2mm}
\end{figure}

\vspace{1mm}
\noindent
\textbf{Auxiliary loss weight $\alpha$.}
The weight $\alpha$ helps ensure balanced training between the two modules. We conducted a sensitivity analysis by varying $\alpha$ over a range of values: [0.05, 0.1, 0.2, 0.5, 1], as shown in Figure~\ref{fig:hyper_aux}(a) and (b) for the Beauty dataset. Excessively small ($\alpha$ < 0.05) or large ($\alpha$ > 0.5) values led to performance degradation. MUFFIN exhibited stable and high performance when $\alpha$ was set within the range of 0.05 to 0.1 for most datasets. This sweet spot allows the auxiliary loss to provide sufficient guidance without overwhelming the primary objective.


\begin{figure}
\begin{tabular}{cc}
\includegraphics[height=3.0cm]{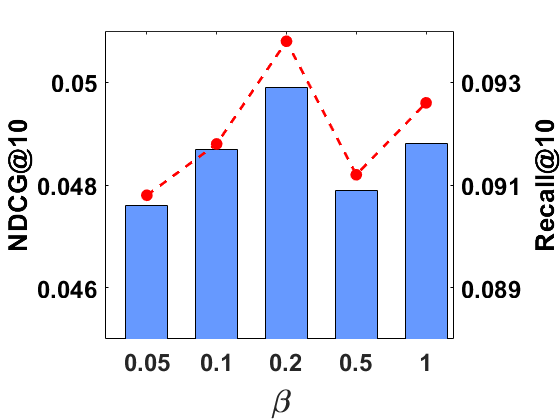} &
\includegraphics[height=3.0cm]{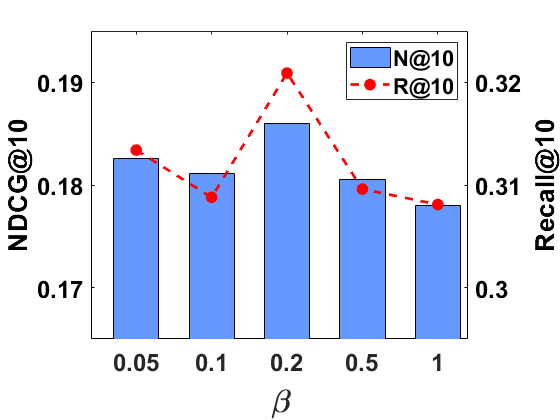} \\
\multicolumn{1}{c}{(a) Beauty} &\multicolumn{1}{c}{(b) ML-1M}
\end{tabular}
\vspace{-3mm}
\caption{Accuracy of MUFFIN over varying load balancing loss weight $\beta$ on Beauty and ML-1M. }\label{fig:hyper_bal}
\vspace{-2mm}
\end{figure}

\vspace{1mm}
\noindent
\textbf{Load balancing loss weight $\beta$.}
The weight $\beta$ controls the balance across frequency bands. Figure~\ref{fig:hyper_bal}(a) and (b) show its impact when varied from [0.05, 0.1, 0.2, 0.5, 1]. Too small $\beta<0.05$ causes probability concentration on specific bands, while too large $\beta>1$ degrades performance by forcing uniform distribution across frequency bands. Most datasets showed the best performance at $\beta=0.2 \sim 1$. This range ensures that MUFFIN utilizes diverse frequency components while focusing on the informative bands for each user.

\vspace{-2mm}
\section{Related Work}

This section reviews existing SR models into two categories: \emph{time-domain models} and \emph{frequency-domain models}.

\vspace{1mm}
\noindent
\textbf{Time-domain models.} Early SR models relied on heuristic-based approaches such as K-nearest neighbor (KNN) methods~\cite{JannachL17, GargGMVS19}, which were limited in modeling complex, long-range dependencies due to their reliance on simple interaction co-occurrences. The emergence of deep learning~\cite{HeLZNHC17} significantly advanced the field, enabling more sophisticated modeling of user behavior through neural architectures. Various neural encoders have been explored, including Convolutional Neural Networks (CNNs)~\cite{TangW18caser}, Recurrent Neural Networks (RNNs)~\cite{LiRCRLM17NARM, HidasiKBT15GRU4Rec}, Graph Neural Networks (GNNs)~\cite{WuT0WXT19SRGNN, GuptaGMVS19NISER}, and Transformers~\cite{vaswani2017attention, KangM18SASRec, SunLWPLOJ19BERT4Rec}. Among them, SASRec~\cite{KangM18SASRec} introduced self-attention mechanisms for modeling item-item dependencies, while BERT4Rec~\cite{SunLWPLOJ19BERT4Rec} leveraged bidirectional Transformers to enhance contextual understanding. In parallel, contrastive learning has emerged as a powerful tool for improving sequence representation learning. For instance, CL4SRec~\cite{XieSLWGZDC22CL4SRec} utilized data augmentation strategies to create robust contrastive pairs, and DuoRec~\cite{QiuHYW22DuoRec} introduced supervised augmentation tasks to enrich training signals. Despite these advancements, time-domain models often struggle to capture periodic patterns common in user behavior sequences.

%Models like SASRec~\cite{KangM18SASRec} pioneered the use of self-attention mechanisms to capture item-item relationships, while BERT4Rec~\cite{SunLWPLOJ19BERT4Rec} adopted bidirectional encoding for better context understanding. Beyond these architectural advancements, recent works have leveraged contrastive learning to enhance sequence representations. CL4SRec~\cite{XieSLWGZDC22CL4SRec} employed data-based augmentation techniques to learn more robust representations, while DuoRec~\cite{QiuHYW22DuoRec} introduced additional supervised augmentation. However, these time-domain SR models struggle to effectively capture periodic behavior patterns.

\vspace{1mm}
\noindent
\textbf{Frequency-domain models.} Recent research has turned to the frequency domain for sequence modeling. FMLP-Rec~\cite{ZhouYZW22FMLPRec} first introduced frequency-based filtering using MLPs to uncover periodic patterns in user-item interactions. Building on this idea, SLIME4Rec~\cite{DuYZF0LS023SLIME4Rec} and FEARec~\cite{DuYZQZ0LS23FEARec} proposed advanced architectures with layered frequency ramp structures and frequency-aware attention mechanisms. Additionally, some studies~\cite{zhang2023contrastive, yang2024adaptive} have incorporated Fourier-based data augmentation for contrastive learning, further enhancing representation robustness. BSARec~\cite{aaai/Shin0WP24BSARec} introduced fine-grained frequency adjustment as an inductive bias in self-attention layers, aiming to better capture subtle sequential signals. Most recently, some studies have also explored frequency filtering to extract users' distinctive information or filter out noisy components~\cite{zhang2025frequency,kim2025diff}. While these frequency-domain models lack user-specific adaptability across users, MUFFIN dynamically identifies diverse behavioral patterns by using dual complementary modules with user-adaptive filters.
% While frequency-domain models excel at modeling periodic behaviors, they struggle to adapt to user-specific preferences, limiting their generalization in diverse behavioral contexts. Our work addresses this problem by using dual user-adaptive filtering modules.

%Recently, researchers have begun exploring frequency domain analysis for sequential recommendation. FMLP-Rec~\cite{ZhouYZW22FMLPRec} first introduced frequency-based MLP filtering to capture periodic patterns, while SLIME4Rec~\cite{DuYZF0LS023SLIME4Rec} and FEARec~\cite{DuYZQZ0LS23FEARec} further advanced this direction by proposing a layered frequency ramp structure with contrastive learning. Following this trend of incorporating contrastive learning, several studies leverage the Fourier transform for data augmentation~\cite{zhang2023contrastive, yang2024adaptive}. Recently, BSARec~\cite{aaai/Shin0WP24BSARec} attempted to find out fine-grained sequential patterns and inject it as inductive bias. Despite their effectiveness in capturing periodic patterns, these frequency-domain SR models often lack user-specific adaptability.

\vspace{-3mm}
\section{Conclusion}\label{sec:conclusion}
In this paper, we present \textbf{MUFFIN}, a novel SR model to overcome two major limitations of existing frequency-domain models: limited frequency band coverage and lack of personalized frequency filtering. MUFFIN adopts a dual filtering architecture, comprising GFM and LFM, to effectively exploit the full frequency spectrum. The UAF also enables user-specific frequency filtering for both filtering modules. Extensive experiments demonstrate that MUFFIN consistently outperforms state-of-the-art baselines. Our in-depth analysis confirms that MUFFIN exhibits variability in filters across users. Future work will explore multi-domain scenarios that exhibit more dynamic user behavior patterns and richer frequency characteristics.

\bibliographystyle{ACM-Reference-Format}
\balance
\bibliography{reference}

\end{document}